\documentclass{appolb} 
\usepackage{epsfig}  

\newcommand{\bea}{\begin{eqnarray}}
\newcommand{\eea}{\end{eqnarray}}
 
\newcommand{\ra}{\rightarrow}

\newcommand{\lsim}{\raisebox{-0.13cm}{~\shortstack{$<$\\[-0.07cm] $\sim$}}~}
\newcommand{\gsim}{\raisebox{-0.13cm}{~\shortstack{$>$\\[-0.07cm] $\sim$}}~}
\def\ifmath#1{\relax\ifmmode #1\else $#1$\fi}

\def\ls#1{\ifmath{_{\lower1.5pt\hbox{$\scriptstyle #1$}}}}
\def\lsup#1{^{\lower 3pt\hbox{$\scriptstyle#1$}}}

\def\Eq#1{Eq.~(\ref{#1})}

\newcommand{\bsllp}{B_s^0 \rightarrow \ell^+ \ell'^{-}}

\newcommand{\cbsdllp}{\mathcal{B}(B_{s,d}^0 \rightarrow \ell^+ \ell'^{-})}
\newcommand{\bsdllp}{B_{s,d}^0 \rightarrow \ell^+ \ell'^{-}}
\newcommand{\bsdmumu}{B_{s,d}^0 \rightarrow \mu^+ \mu^{-}}

\newcommand{\cbsmumu}{\mathcal{B}(B_s^0 \rightarrow \mu^+ \mu^{-})}

\newcommand{\cbsdmumu}{\mathcal{B}(B_{s,d}^0 \rightarrow \mu^+ \mu^{-})}
\newcommand{\bs}{B_s^0 \rightarrow \mu^+ \mu^{-}}
\newcommand{\bd}{B_d^0 \rightarrow \mu^+ \mu^{-}}

\begin{document}

%


\title{MSSM Predictions for $B^0 \rightarrow \mu^+ \mu^-$ at Tevatron
  and LHC\thanks{Presented at the FlaviaNet Topical Workshop, 23-27
    July 2009, Kazimierz, Poland.}}
\author{Janusz Rosiek
\address{Institute of Theoretical Physics, University of Warsaw}
}

\maketitle

\begin{abstract}
During the last few years the Tevatron has improved the bounds on rare
$B$-meson decays into two leptons.  Sensitivity to this decay is also
one of the benchmark goals for LHCb performance.
We compute the complete 1-loop MSSM contribution to $\bsdmumu$ and
study the predictions for arbitrary flavour mixing parameters.  We
discuss the possibility of both enhancing and suppressing the
branching ratios relative to their SM expectations.
We find that there are ``cancellation regions'' in parameter space
where the branching ratio is suppressed well below the SM
expectation, making it effectively invisible to the LHC.
\end{abstract}

  
\section{Introduction}
\label{sec:intro}

One of the most promising signals for new physics at the LHC is the
rare decay $B^0_{s} \rightarrow \mu^+\mu^-$.  The decay is strongly
suppressed but especially `clean' because its final state is easily
tagged and its only hadronic uncertainties come from the hadronic
decay constant $f_{B_{s}}$~\cite{Buchalla:1993bv}.  The LHC will be
the first experiment to be able to probe this decay channel down to
its Standard Model (SM) branching ratio.  In particular the LHCb will
be able to directly probe the SM predictions at $3\sigma$ ($5\sigma$)
significance with 2 $\mathrm{fb}^{-1}$ (6 $\mathrm{fb}^{-1}$) of data,
or after about 1 year (3 years) of design
luminosity~\cite{arXiv:0710.5056}.  It is not clear whether LHC can
reach the SM expectation for $\bd$.

The current experimental status and the SM predictions for the
branching ratios $\bsdmumu$ are:\\
\begin{tabular}{llll}\hline \hline
\textbf{Channel} & \textbf{Expt.} & \textbf{Bound (90\% CL)} &
\textbf{SM Prediction} \\
\hline $B^0_s \to \mu^+ \mu^-$ & CDF II~\cite{:2007kv} & $<4.7\times
10^{-8}$ & $(4.8\pm 1.3) \times 10^{-9}$ \\ 
$B^0_d \to \mu^+ \mu^-$ & CDF II~\cite{:2007kv} & $<1.5\times 10^{-8}$
& $(1.4\pm 0.4) \times 10^{-10}$ \\ 
\hline 
\end{tabular}\\

At the dawn of the LHC era, it is important to understand the possible
contributions of new physics to the $\bsdmumu$ decays. They could be
particularly large in the Minimal Supersymmetric Standard Model
(MSSM).  Under the assumption of large $\tan\beta$ and Minimal Flavour
Violation (MFV, flavour violation given by CKM matrix only), the
branching ratio for $B^0_s \rightarrow \mu^+\mu^-$ is dominated by the
Higgs penguin mode and can be significantly enhanced over the SM
expectation, as can be seen from the approximate formulae (for
detailed study focusing on the resummation of $\tan\beta$
see~\cite{Buras:2002vd}):
\vspace{-2mm}
\bea
{\cal B}(B^0_s\rightarrow \mu^+\mu^-) &\approx& 5 \cdot 10^{-7}
\left(\frac{\tan\beta}{50}\right)^6\left(\frac{300 \hspace{.2cm}
  \mathrm{GeV}}{M_A}\right)^4 .
\label{eq:tanbeta}
\eea
\vskip -2mm
With the upcoming new experimental probes of $\cbsdmumu$, it is
important to perform a full calculation of this decay rate without
\textit{a priori} assumptions on the pattern of flavour mixing.  In
particular, in the region of low $\tan \beta$, the interference of
box, $Z$- and Higgs-penguin diagrams could conceivably lead to both
enhancement (even visible at Tevatron) or a cancellation that would
suppress the branching ratio below the SM prediction (making it
invisible at the LHC).  The status of this decay could become an
important factor for planned LHCb upgrades and determining whether
they should focus on increasing sensitivity to $\bs$ or instead reach
for the smaller branching ratio of $\bd$. The relevant calculations
have been presented in details in~\cite{DRT}, we stress here the most
important results.

\section{Effective Operators and Branching Ratios}
\label{sec:operators}

The effective Hamiltonian for the quarks-to-leptons transition $q^{I}
q^{J} \rightarrow \ell^{+K} \ell^{-L}$, with $q^{1}\equiv d,
q^{2}\equiv s, q^{3}\equiv b$ and $\ell^{1}\equiv e, \ell^{2}\equiv
\mu, \ell^{3}\equiv \tau$, reads:
\vspace{-3mm}
\bea
{\cal H} \ = \ \frac{1}{(4\pi)^2}\sum_{X,Y=L,R}\biggl ( C_{VXY}
\mathcal{O}_{VXY} \ + \ C_{SXY} \mathcal{O}_{SXY} \ + \ C_{TX}
\mathcal{O}_{TX} \biggr ) , 
\label{ham}
\eea
\vskip -3mm
\noindent where flavour and colour indices have been suppressed for
brevity.  The (V)ector and (S)calar operators are respectively given
by
\vspace{-3mm}
\bea
\mathcal{O}_{VXY}^{IJKL} \ &=& \ (\overline{q^{J}}  \gamma^{\mu}
P_{X} q^{I} ) (\overline{\ell^{L}} \gamma_{\mu} P_{Y} 
\ell^{K} ) , \nonumber \\
\mathcal{O}_{SXY}^{IJKL} \ &=& \ (\overline{q^{J}}  P_{X}  q^{I} )
(\overline{\ell^{L}}  P_{Y}  \ell^{K} ) . \label{effops}
\eea
\vskip -3mm
The explicit form of the Wilson coefficients in the MSSM is given
in~\cite{DRT}. The (T)ensor operator contributions and photon penguin
contribution to $\bsllp$ vanishes in matrix element calculations.  We
do not consider the very large $\tan\beta\gsim30$ scenario, thus no
resummation of higher orders in $\tan\beta$ is necessary.  The matrix
element for $\bsdllp$ decay is:
\vspace{-3mm}
\bea\label{me}
\mathcal{M} \ =\ F_{S}  \overline{\ell}\ell \ + \ F_{P}
\overline{\ell}\gamma_{5} \ell \ + \ F_{V} p^{\mu}
\overline{\ell}\gamma_{\mu} \ell \ + \ F_{A} p^{\mu}
\overline{\ell}\gamma_{\mu} \gamma_{5}\ell ,
\eea
\vskip -3mm
\noindent where the $\ell$s correspond to external lepton spinors.
The (S)calar, (P)seudo\-sca\-lar, (V)ector and (A)xial-vector form
factors in~\Eq{me} are given by
\vspace{-3mm}
\bea
F_{S} \ &=& \ \frac{i}{4}  \frac{M_{B_{s(d)}}^{2}
  f_{B_{s(d)}}}{m_{b}+m_{s(d)}}  (C_{SLL} + C_{SLR} - C_{SRR}
-C_{SRL} ) , \label{fs} \\
F_{P} \ &=& \ \frac{i}{4}  \frac{M_{B_{s(d)}}^{2}
  f_{B_{s(d)}}}{m_{b}+m_{s(d)}}  (-C_{SLL} + C_{SLR} - C_{SRR} +
C_{SRL} ) , \label{fp} \\
F_{V} \ &=& \ -\frac{i}{4} f_{B_{s(d)}}  (C_{VLL} + C_{VLR} -
C_{VRR} - C_{VRL} ) , \label{fv} \\
F_{A} \ &=& \ -\frac{i}{4} f_{B_{s(d)}}  (- C_{VLL} + C_{VLR} -
C_{VRR} + C_{VRL} ) .  \label{fa}
\eea
\vskip -2mm
The general expression for $\cbsdllp$ is rather
complicated~\cite{DRT}.  For the most important $\bsdmumu$ decays it
simplifies greatly and reads approximately as ($q=s,d$)
\vspace{-2mm}
\bea
\mathcal{B}(B_q^0 \ra \mu^-  \mu^+)
\ \approx\ \frac{\tau_{B_q}M_{B_q}}{8\pi} \left ( |F_S|^2 \ + \ |F_P
+ 2 m_\mu F_A |^2 \right )  ,\label{brr}
\eea
\vskip -2mm
\noindent where $\tau_{B_q}$ is the lifetime of $B_q$ meson and we
have taken the limit $\frac{m_\mu}{M_{B_q}} \to 0$.
\vspace{-3mm}

\section{Numerical Analysis of $B_{s,d}\ra \mu^+\mu^-$}
\label{sec:analysis}

We may distinguish two possible scenarios for the relative size of the
MSSM contributions to the right-hand side of~\Eq{brr}:\\[1mm]
1. {\sl Higgs penguin domination or large $\tan\beta \gsim 10$}.  In
this regime one can expect large $F_S,F_P$ and an enhancement of the
branching ratios as in \Eq{eq:tanbeta} (barring some possible GIM-type
cancellations leading to $F_{S,P}^{\rm SUSY} \approx 0$
~\cite{Dedes:2002er}).\\[1mm]
2. {\sl Comparable Box, $Z$- and Higgs-penguin contributions or low
  $\tan\beta \lsim 10$}.  In this case the supersymmetric
Higgs-mediated form factors $F_{S,P}$ may become comparable to or even
smaller than $F_{A}$.  Either an enhancement or a suppression of the
branching ratios is possible depending on the particular choice of
MSSM parameters.

An enhancement of the branching ratios occurs generically in most of
the MSSM parameter space.  It is a bit trickier to suppress the
branching ratios below their SM predictions. This is the case we would
like to investigate further.  We would like to find the minima of
$\cbsdmumu$, i.e. the minima of \Eq{brr}. We distinguish between two
cases:
\vspace{-2mm}
\bea
&a)&F_{P} +  2 m_\ell F_A \approx 0 \qquad  
\mathrm{and}\qquad  F_{P} \gg F_{S} , \label{c1}\\
&a)&|F_S| \approx  |F_P| \approx |F_A| \approx 0 .   \label{c2}
\eea
\vskip -2mm
In the case a), the pseudoscalar and axial contributions cancel while
the scalar contribution is negligible.  The case b), happens when
Higgs contributions are negligible compared to the axial contribution
(i.e. low $\tan\beta$ and large $M_{A}$) and $F_{A}$ becomes small due
to cancellations among the $C_{VXY}$ coefficients in~\Eq{fa}. Our
numerical analysis shows that both can-

{\small
\begin{tabular}{lcccc}
{\bf Table 1}\\ 
\hline \hline
Parameter & Symbol & Min & Max & Step \\ \hline
Ratio of Higgs vevs & $\tan\beta$ & 2 & 30 & varied \\
CKM phase & $\gamma$ & $0$& $\pi$ & $\pi/25$ \\
CP-odd Higgs mass & $M_{A}$ & 100 & 500 & 200 \\ 
SUSY Higgs mixing & $\mu$ & -450 & 450 & 300 \\ 
$SU(2)$ gaugino mass & $M_{2}$ & 100 & 500 & 200 \\ 
Gluino mass & $M_{3}$ &$3M_{2}$&$3M_{2}$& 0 \\ 
SUSY scale & $M_{\mathrm{SUSY}}$ & 500 & 1000 & 500 \\ 
Slepton Masses & $M_{\tilde{\ell}}$& $M_{\mathrm{SUSY}}/3$ &
$M_{\mathrm{SUSY}}/3$ & 0 \\
Left top squark mass & $M_{\tilde{Q}_{L}}$ & 200 & 500& 300 \\ 
Right bottom squark mass & $M_{\tilde{b}_{R}}$ & 200& 500& 300 \\
Right top squark mass & $M_{\tilde{t}_{R}}$ & 150 & 300 & 150 \\
Mass insertion & $\delta_{dLL}^{13}$, $\delta_{dLL}^{23}$ & -1& 1& 1/10 \\
Mass insertion & $\delta_{dLR}^{13}$, $\delta_{dLR}^{23}$ & -0.1& 0.1& 1/100 \\ \hline
\end{tabular} \\
}

\noindent cellations are possible but require a certain amount of fine
tuning once constraints from other FCNC measurements are imposed.

To quantitatively study the effects mentioned above we perform a scan
over the MSSM parameters, not restricted to MFV scenario.  Flavour
violation is parametrised by the ``mass insertions'', defined as in
\cite{Gabbiani:1996hi, Misiak:1997ei}
\vspace{-2mm}
\bea
\delta^{IJ}_{QXY} &=& \frac{(M^2_{Q})^{IJ}_{XY}}{\sqrt{
    (M^2_{Q})^{IJ}_{XX} (M^2_{Q})^{IJ}_{YY} }} ,
\label{eq:phys:massinsert}
\eea
\vskip -2mm
\noindent where $I,J$ are squark flavours, $X,Y$ denote field
chirality, and $Q$ indicates the up or down squark sector. Note that
we use the $\delta$'s for presentation only, not as expansion
parameters - we numerically diagonalize all mass matrices.

The ranges of variation over MSSM parameters are shown in Table 1,
where ``SUSY scale'' denotes the common mass parameter for the first
two squark generations, $\tan\beta$ takes on values
$(2,4,6,8,10,13,16,19,22,25,30)$, $\delta_{dLL}^{IJ}$,
$\delta_{dLR}^{IJ},\mu, M_2$ are taken to be real and the trilinear
soft couplings are set to $A_{t}=A_{b}=M_{\tilde{Q}_{L}}$ and
$A_{\tilde{\tau}}=M_{\tilde{L}}$.

To realistically estimate the allowed range for $\cbsdmumu$, one must
account for the experimental constraints from other rare decays.  For
that, we use the library of numerical codes developed in the framework
of the general MSSM in~\cite{Buras:2002vd, DRT, Misiak:1997ei,
  codelib, Buras:2004qb} and take into account the set of observables
listed in Table~2\footnote{For Higgs mass we use LEP data~\cite{LEP}
  $m_h\geq 92.8-114$ GeV depending on $\sin^2(\alpha-\beta)$.)}.  In
our scan, for the quantities in Table~2 we require (depending if the
experimental result or only the upper bound is known)
\vspace{-2mm}
\bea
\mathrm{or}\qquad
\begin{array}{c}
|Q^{exp} - Q^{th}| \leq 3\Delta Q^{exp} + q |Q^{th}|,\\ \\
(1+q)|Q^{th}|\leq Q^{exp}.
\end{array}
\label{eq:xacc}
\eea
%

{\small
\begin{tabular}{lcc}
{\bf Table 2}\\
\hline \hline
Quantity &  Current Measurement & Experimental Error \\ \hline
$m_{\chi^0_1}$ & $>$ 46 ~{\rm GeV} & \\
$m_{\chi^\pm_1}$ & $>$ 94 ~{\rm GeV} & \\
$m_{\tilde b}$ &  $>$ 89 ~{\rm GeV} & \\
$m_{\tilde t}$ & $>$ 95.7~{\rm GeV} & \\
$m_{h}$ & $>$ 92.8 ~{\rm GeV} & \\
$|\epsilon_{K}|$ & $2.232 \cdot 10^{-3}$ & $0.007 \cdot 10^{-3}$ \\
$|\Delta M_{K}|$ & $3.483 \cdot 10^{-15}$ & $0.006 \cdot 10^{-15}$ \\
$|\Delta M_{D}|$ & $ <0.46 \cdot 10^{-13}$ &  \\
$\Delta M_{B_d}$ & $3.337 \cdot 10^{-13}~{\rm GeV}$ & $0.033 \cdot
10^{-13}~{\rm GeV}$ \\
$\Delta M_{B_s}$ & $116.96 \cdot 10^{-13}~{\rm GeV}$ & $0.79 \cdot
10^{-13}~{\rm GeV}$ \\
Br($B\rightarrow X_{s} \gamma$) & $3.34\cdot 10^{-4}$ & $0.38 \cdot
10^{-4}$ \\
Br($K_{L}\rightarrow \pi^{0} \nu \bar{\nu}$) & $<1.5\cdot 10^{-10}$
&\\
Br($K^+\rightarrow \pi^+ \nu \bar{\nu}$) & $1.5\cdot 10^{-10}$ &
$1.3\cdot 10^{-10}$\\
Electron EDM & $<0.07\cdot 10^{-26}$ &\\
Neutron EDM & $<0.63\cdot 10^{-25}$ &\\
\hline \end{tabular}
}\\[1mm]

\noindent $3\Delta Q^{exp}$ and $|q Q^{th}|$ in \Eq{eq:xacc} represent
the $3\sigma$ experimental error and the theoretical error
respectively.  The latter differs from quantity to quantity and is
usually smaller than the value $q=50\%$ which we assume generically in
all calculations.  The increased ``theoretical error'' is used to
account for the limited density of a numerical scan, simultaneously
avoiding strong fine tuning between MSSM parameters (for detailed
discussion see~\cite{Buras:2004qb}).

\begin{figure}[htb]
  \begin{center} \begin{tabular}{cc}
  \epsfig{file=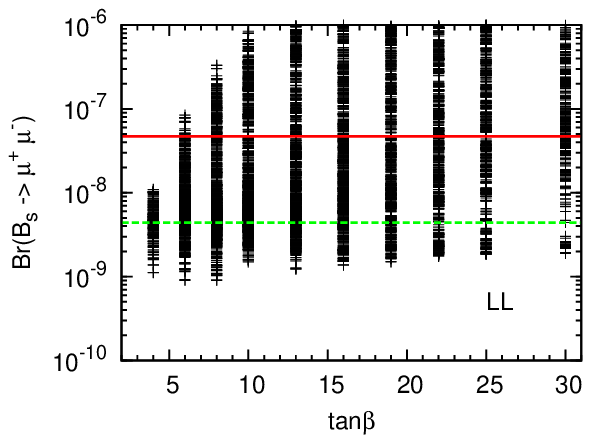,width=0.48\linewidth} &
  \epsfig{file=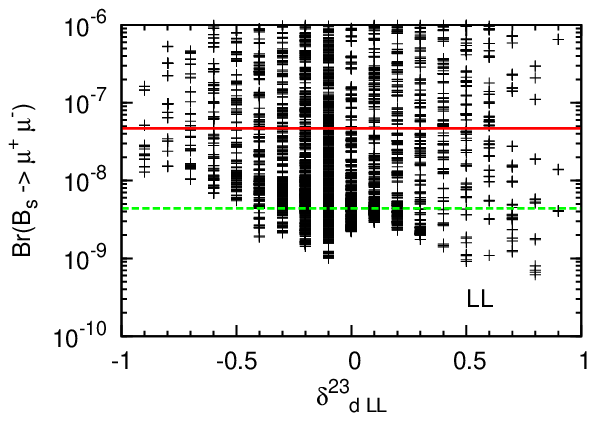,width=0.48\linewidth} \\
  \epsfig{file=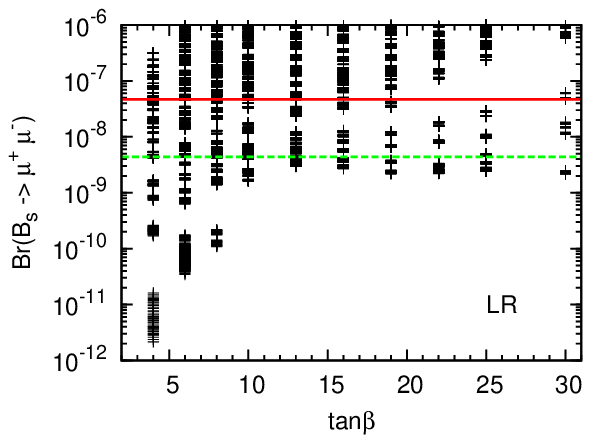,width=0.48\linewidth} &
  \epsfig{file=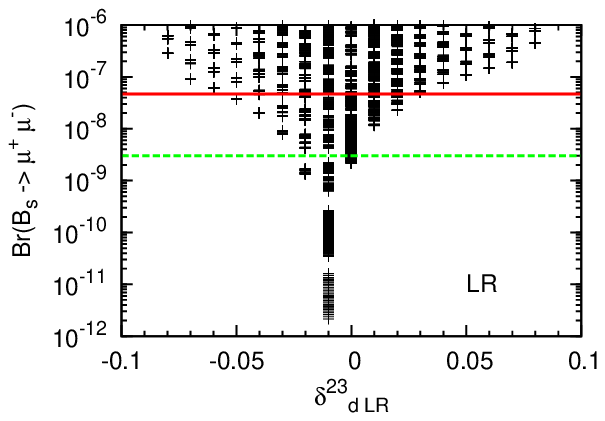,width=0.48\linewidth}
  \end{tabular}
\vspace{-3mm}
\caption{ Upper panel: Predictions for ${\mathcal{B}}(B_s \to \mu^+
  \mu^-)$ versus $\tan\beta$ (left) and $\delta_{d  LL}^{23}$
  (right) from the scan of MSSM parameters.  The upper solid line
  shows the current upper bound from the Tevatron and the lower dashed
  line the SM expectation.  Lower panel: Similar to the upper panel
  but with $\delta_{d  LR}^{23}$ varied.}
\vspace{-10mm}
\label{fig2}
  \end{center}
\end{figure}
Fig.~\ref{fig2} shows the predictions for $\cbsmumu$ over a general
scan of 20 million points in parameter space, including the bounds of
Table~2.  We vary $\delta_{d LL}^{23}$ (upper panel) and $\delta_{d
  LR}^{23}$ (lower panel) one at a time while setting the other to
zero (results are also weakly sensitive to other $\delta$'s).
When $\delta_{d LL}^{23}$ is varied in the range $[-1, 1]$, we find
$\cbsmumu_{min} \approx 10^{-9}$.  This minimum is almost independent
of $\tan\beta$ but depends on the magnitude of $|\delta_{d
  LL}^{23}|$. The upper bound set by CDF, depicted as a solid red
line, can be attained even with very low values of $\tan\beta$.

\begin{figure}
  \begin{center}
    \begin{tabular}{cc}
      \epsfig{file=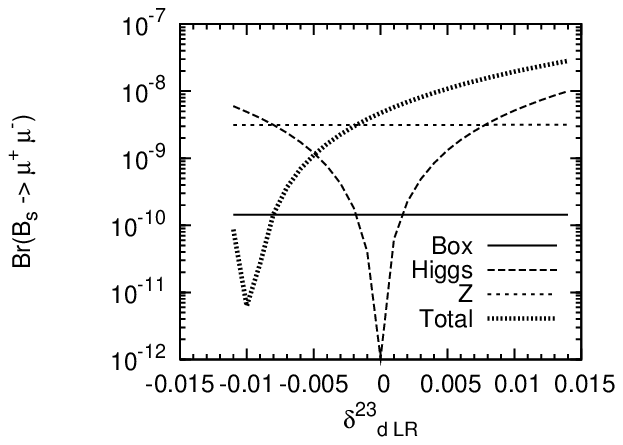,width=0.48\linewidth} &
      \epsfig{file=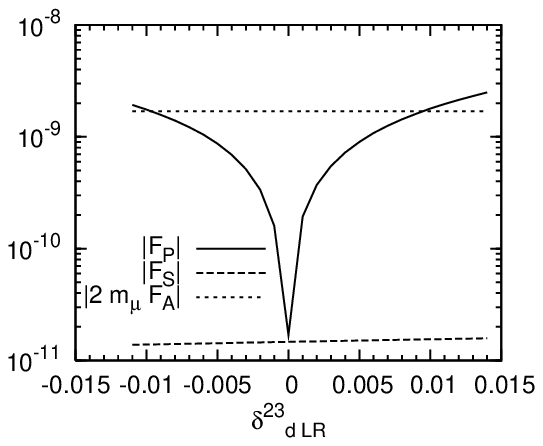,width=0.48\linewidth}
    \end{tabular}
\caption{Contributions to $\cbsdmumu$ for the parameters in \Eq{poi}
  versus $\delta^{23}_{d LR}$. Left: Contributions from the various
  diagram classes.  Right: Magnitude of the form factors appearing in
  Eqs.~(\ref{fs}--\ref{fa}).}
\label{fig3}
\vspace{-10mm}
\end{center}
\end{figure}

More interesting is the case when $\delta_{d LR}^{23}$ is varied in
the range $[-0.1,0.1]$. We find a narrow cancellation region around
$\delta_{d LR}^{23}\approx-0.01$ and $\tan\beta \lsim 10$ where
$\cbsmumu_{min} \approx 10^{-12}$ (lower right panel). This is three
orders of magnitude lower than the SM prediction, making it
effectively unobservable at the LHC.
In order to better understand cancellation region we study a
representative point with a very low branching ratio (all masses in
GeV):
\vspace{-2mm}
\bea
\tan\beta = 4, \quad 
M_{A} = 300, \quad \mu =-450, \quad M_{2}=100, \quad M_{3}=300 ,
\nonumber \\
\mathrm{SUSY ~scale}= 400, \quad M_{\tilde{t}_R}= 150, \quad 
A_{t,b}=M_{\tilde{t}_L}=
M_{\tilde{b}_{(L,R)}}=600 .\label{poi}
\eea
\vskip -3mm
\noindent The `Box', `Higgs' and `$Z$' lines in Fig.~\ref{fig3}
indicate the value of $\cbsmumu$ given by only the listed contribution
with all others set to zero.  In the cancellation region the box
contribution is negligible while the Higgs- and $Z$-penguin magnitudes
are comparable. Thus the latter should cancel, which we illustrate
individually plotting the absolute values of the form factors
$F_{S,P}$ and $2 m_{\mu} F_{A}$ of Eqs.~(\ref{fs}--\ref{fa}) in the
right panel of Fig.~\ref{fig3}.  At the minimum point of the total
branching ratio $|F_{P}|$ is approximately equal to $|2 m_{\mu}
F_{A}|$ and $|F_{S}|$ is negligibly small.
This can be explained from the form of Eqs.~(\ref{fs}) and (\ref{fp}).
If one assumes $\delta^{23}_{d LR}=\left(\delta^{32}_{d
  LR}\right)^\star$, then $C_{SLR}$ and $C_{SRL}$, the two Wilson
coefficients most sensitive to the variation of $\delta^{23}_{dLR}$,
have similar sizes and opposite sign, interfering destructively in the
amplitude.

We performed similar scan also for the $B_d$ meson decay, $B_d \to
\mu^+ \mu^-$, varying $\delta_{d LL}^{13}$ or $\delta_{d LR}^{13}$
instead of $\delta_{d LL}^{23}$ or $\delta_{d LR}^{23}$ along with the
other SUSY parameters.
For both cases there exist points where ${\mathcal{B}}(B_d \to \mu^+
\mu^-)$ is reduced by an order of magnitude relative to the SM. These
points are more sensitive to low $\tan\beta$ in the `LL' case and fall
into the case of \Eq{c2}.  On the opposite, the ratio
$R={\mathcal{B}}(B_d \to \mu^+ \mu^-)/{\mathcal{B}}(B_s \to \mu^+
\mu^-)$, which in the SM is fixed to $R \approx |V_{td}/V_{ts}|^2 \leq
0.03$, in the MSSM can be enhanced by a factor of 10 even for small
values of $\delta^{13}_{d LL}$ or $\delta^{13}_{d LR}$.  This suggests
that collider searches for ${\mathcal{B}}(B_d \to \mu^+ \mu^-)$ are as
important as those for ${\mathcal{B}}(B_s \to \mu^+ \mu^-)$.

Our results lead also to bounds on $\delta$ parameters. As can be seen
from Figs~\ref{fig2} and scan results for ${\mathcal{B}}(B_d \to \mu^+
\mu^-)$, $\delta^{23}_{dLL},\delta^{13}_{dLL}$ are weakly constrained,
they can take on values up to $\approx 0.9$ and still pass the
constraints in Table~2, though points beyond $0.3$ are less
dense. Bounds in the 'LR' sector are tighter, $\delta^{23}_{d
  LR},\delta^{13}_{d LR}\lsim 0.08$.  \vspace{-5mm}

\section{Conclusions}
\label{sec:conclusion}

We have discussed results of a complete, 1-loop calculation of the
branching ratios for the rare decay modes $\bsdmumu$ without resorting
to the limits of large $\tan\beta$ and MFV scenario and performed a
numerical exploration of the MSSM parameter space. We find that there
exist cancellation regions where the contribution of diagrams with
supersymmetric particles interferes destructively with purely SM
diagrams, thus allowing the branching ratio to be significantly
smaller than the SM prediction.  We identify possible mechanisms of
such cancellations and explain why they can occur for certain regions
of parameter space.  Such effects may effectively hide the dimuon
$B^0_s$ decay mode from the LHCb even though it is supposed to be one
of the experiment's benchmark modes.  We have also shown that, barring
the cancellations mentioned above, supersymmetric contributions in the
general MSSM typically tend to enhance the branching ratio for
$\bsdmumu$ even for moderate values of $\tan\beta\lsim 10$ so that an
experimental measurement close to the SM prediction would put strong
bounds on the size of allowed flavour violation in the squark sector.
Finally, we show that the $B^0_d\ra \mu^+\mu^-$ decay can also be
either suppressed or enhanced compared to its SM expectation, leading
in some cases to a situation where the rate of the $B^0_d$ decay is
larger then that of the $B^0_s$.
\vspace{1mm}

\noindent {\bf Acknowledgements.}  The paper was supported in part by
the RTN European Programme, MRTN-CT-2006-035505 (HEPTOOLS, Tools and
Precision Calculations for Physics Discoveries at Colliders) and by
the Polish Ministry of Science and Higher Education Grant
N~N202~230337.

\end{document}